\documentstyle[12pt,amssymb,amsmath]{article}
\newtheorem{theorem}{Theorem}[section]
\newtheorem{lemma}{Lemma}[section]

\newcommand{\p}{\partial}
\newcommand{\la}{\lambda}

\begin{document}

\title{ {\bf Two binary Darboux transformations for the KdV hierarchy 
with self-consistent sources  }  }
\author{ {\bf  Yunbo Zeng\dag 
	\hspace{1cm} Wen-Xiu Ma\ddag \hspace{1cm} Yijun Shao\dag\hspace{1cm}} \\
    {\small {\it \dag
	Department of Mathematical Sciences, Tsinghua University,
	Beijing 100084, China}} \\
    {\small {\it \ddag
	Department of Mathematics, City University of Hong Kong, 
	Kowloon, Hong Kong, China}}  }  
\date{}
\maketitle
\renewcommand{\theequation}{\arabic{section}.\arabic{equation}}

\begin {abstract}
Two binary (integral type) Darboux transformations for the KdV hierarchy 
with self-consistent sources are proposed. In contrast with the Darboux 
transformation for the KdV hierarchy, one of the two binary Darboux 
transformations provides non auto-B\"{a}cklund transformation between 
two n-th KdV equations with self-consistent sources with different degrees. 
The formula for the  m-times repeated binary Darboux transformations 
are presented.  This enables us to construct the N-soliton solution 
for the KdV hierarchy with self-consistent sources.
\end{abstract}

\hskip\parindent

 
\hskip\parindent

\section{Introduction}
\setcounter{equation}{0}
\hskip\parindent
The soliton equations with self-consistent sources  have important
physical applications
(see \cite{mel89a}-\cite{Shch96}), for example, the KdV equation 
with self-consistent source describes the interaction of long and 
short capillary-gravity waves \cite {Lati90}. There are some ways 
to derive the integrable nonlinear evolution equations with 
self-consistent sources \cite{mel89a,mel89b,mel90a, mel90,Leon89}.
 In recent years soliton equations with self-consistent sources (SESCS) 
were studied based on the constrained flows of soliton equations which 
are just the stationary  equations of SESCSs \cite{Anto92}-\cite{ma96}. 
 Since the Lax representations for the 
constrained flows of soliton equations can always be deduced from 
the adjoint representations of the Lax representations for soliton equations, 
this approach provides
a simple and natural way to derive both the SESCSs and 
their Lax representations \cite{Zeng93,Zeng94,Zeng96}. 
The SESCS is an infinite-dimensional integrable Hamiltonian system 
possessing $t-$type 
Hamiltonian or bi-Hamiltonian formulation \cite{Zeng99} and 
can be solved by the inverse scattering method 
\cite{mel89a,mel89b,mel90a,mel88,mel92,Zeng00}.

The Darboux transformation is a power tool for solving soliton equations 
(see \cite{salle} for a review). The Darboux transformation for KdV hierarchy 
was widely studied (see, for example, \cite{salle}-\cite{chen}). 
In the present paper we will generalize these results to the KdV hierarchy 
with self-consistent sources. We construct one Darboux transformation and 
two binary (integral type) Darboux transformations for the KdV hierarchy 
with self-consistent sources. 
The Darboux transformations usually present  auto-B\"{a}cklund transformations 
for soliton equations. In contrast with the case of soliton equations,  
one binary Darboux transformation in our case is proved to 
be a non auto-B\"{a}cklund transformation between two n-th
KdV equations with self-consistent sources with different degrees. 
This provides an interesting example for constructing 
non auto-B\"{a}cklund transformations by means of Darboux transformations. 
Furthermore we present the formula for the  m-times repeated 
binary Darboux transformations  and  
construct the N-soliton solution for the KdV hierarchy with self-consistent sources.

The paper is organized as follows. In the next section we recall 
the KdV hierarchy with self-consistent sources and briefly describe 
how to derive their Lax representation from the adjoint representation of 
the Lax representation for the KdV hierarchy. In section 3, we briefly 
review  Darboux transformations for the KdV hierarchy and present 
two binary Darboux transformations. Based on these results, in section 4,  
we propose one Darboux transformation and two binary Darboux transformations 
for the KdV hierarchy with self-consistent sources  and show that 
the first binary Darboux transformation gives the  auto-B\"{a}cklund 
transformation for the KdV hierarchy with self-consistent sources, 
and the second binary Darboux transformation leads to a non auto-B\"{a}cklund 
transformation relating two n-th KdV equations with self-consistent sources 
with different degrees. Finally in the last section we present the m-times 
repeated binary Darboux transformations and construct the N-soliton 
solution for the n-th KdV equation with self-consistent sources.

\section{The KdV hierarchy with self-consistent sources}
\setcounter{equation}{0}
\hskip\parindent
To make the paper self-contained, we first recall 
the high-order constrained flows of the KdV hierarchy 
and  briefly describe how to derive the Lax representation for 
the KdV hierarchy with 
self-consistent sources.

Consider the Schr\"{o}dinger equation
\begin{equation}
\label{a1}
	\phi_{xx}+(\lambda+u) \phi =0.
\end{equation}
In order to derive the Lax representation for the KdV hierarchy with 
self-consistent sources, we rewrite the
equation (\ref{a1})  in 
the matrix form
\begin{equation}
\label{eqn:kdv-spe}
	\left( \begin{array}{c} 
	\phi \\ \phi_x \end{array} \right)_x=
	U \left( \begin{array}{c}
	\phi \\ \phi_x \end{array} \right),
	\qquad
	U = \left( \begin{array}{cc}
	0 & 1 \\ -\lambda-u & 0 \end{array} \right).
\end{equation}
The adjoint representation of (\ref{eqn:kdv-spe}) reads \cite{newell}
\begin{equation}
\label{eqn:p1}
V_x=[U, V]\equiv UV-VU.
\end{equation}
Set
\begin{equation}
\label{eqn:p2}
	V=\sum_{i=0}^{\infty} \left( \begin{array}{cc}
	a_i & b_i \\ c_i & -a_i \end{array} \right)\lambda^{-i}.
\end{equation}
Eq. (\ref{eqn:p1}) yields
$$a_0=b_0=0,\quad c_0=-1,\quad a_1=0, \quad b_1=1,\quad c_1=-\frac 12u,$$
$$a_2=\frac 14u_x,\quad b_2=-\frac 12u,\quad c_2=\frac 18(u_{xx}+u^2),...,$$
and in general for $k=1,2,...,$
\begin{equation}
\label{eqn:p3}
 a_k=-\frac{1}{2}b_{k,x},\quad
	b_{k+1}=Lb_k=-\frac{1}{2}L^{k-1}u,
	\quad c_k=-\frac{1}{2}b_{k,xx}-b_{k+1}-b_k u,
\end{equation}
where
$$ L=-\frac{1}{4}\p^2-u+\frac{1}{2}\p^{-1}u_x,
	\quad \p=\frac{\partial}{\partial x},
	\quad \p\p^{-1}=\p^{-1}\p =1. $$

Set
\begin{equation}
\label{eqn:p4}
	V^{(n)}=\sum_{i=0}^{n+1} \left( \begin{array}{cc}
	a_i & b_i \\ c_i & -a_i \end{array} \right)\lambda^{n+1-i}
      + \left( \begin{array}{cc}
	0 & 0 \\ b_{n+2} &0 \end{array} \right),
\end{equation}
and take
\begin{equation}
\label{eqn:p5}
	\left( \begin{array}{c} 
	\phi \\ \phi_x \end{array} \right)_{t_n}=
	V^{(n)}(u, \lambda) \left( \begin{array}{c}
	\phi \\ \phi_x \end{array} \right),
\end{equation}
or equivalently
\begin{equation}
\label{a2}
	\phi_{t_n}=A^{(n)}(u, \lambda)\phi,\qquad A^{(n)}(u, \lambda)\equiv
\sum_{i=0}^{n+1} (a_i + b_i\p)\lambda^{n+1-i}.
\end{equation}
Then the compatibility condition of Eqs (\ref{a1}) and (\ref{a2}) or (\ref{eqn:kdv-spe}) 
 and (\ref{eqn:p5}) 
gives rise to the KdV hierarchy
\begin{equation}
\label{eqn:p6}
	u_{t_n}=K_n[u]\equiv \p\frac{\delta H_n}{\delta u}\equiv
	-2b_{n+2,x}, \qquad n=0,1,\cdots,
\end{equation}
where $H_n=\frac {4b_{n+3}}{2n+3}$.
We have
\begin{equation}
\label{eqn:p7}
\frac{\delta \lambda}{\delta u}=\phi^2,\quad\quad L\phi^2=\lambda\phi^2.
\end{equation}

The high-order constrained flows of the KdV hierarchy consist of 
the equations obtained from the spectral problem (\ref{a1}) 
for $N$ distinct $\lambda_j$ and the restriction of 
the variational derivatives for the conserved quantities $H_n$ and $\lambda_j$ \cite{zeng91}

\begin{subequations}
\label{eqn:p8}
\begin{equation}
\label{eqn:p8a}
	D\left[\frac{\delta H_n}{\delta u} -
	2\alpha\sum_{j=1}^N 
	\frac{\delta \lambda_j}{\delta u}\right] \equiv
D\left[-2b_{n+2}-2\alpha\sum_{j=1}^N 
	{\phi_j}^2\right]=0,
\end{equation}
\begin{equation}
\label{eqn:p8b}
	\phi_{j,xx}+(\lambda_j+u)\phi_j=0,\qquad j=1,\cdots,N,
\end{equation}
\end{subequations}
where $n=0,1,\cdots.$  According to Eqs (\ref{eqn:p3}), 
(\ref{eqn:p7}) and (\ref{eqn:p8}), we may define
$$\widetilde a_i=a_i,\quad \widetilde b_i=b_i,\quad
\widetilde c_i=c_i,\quad i=0,1,...,n+1,$$
$$\widetilde b_{n+2+i}=-\alpha \sum_{j=1}^N\lambda^i_j\phi^2_j,
\quad\widetilde a_{n+2+i}=-\frac 12\widetilde b_{n+2+i,x}=
\alpha \sum_{j=1}^N\lambda^i_j\phi_j\phi_{j,x},\quad
i=0,1,2,...,$$
$$\widetilde c_{n+2+i}=-\frac 12\widetilde b_{n+2+i,xx}
-\widetilde b_{n+3+i}-\widetilde b_{n+2+i}u=\alpha \sum_{j=1}^N \lambda^i_j\phi^2_{j,x}.$$
Then the construction of $\widetilde a_i, \widetilde b_i, \widetilde c_i$ ensures that
$$ N^{(n)} 
 =\lambda^{n+1} \sum_{k=0}^{\infty} \left( \begin{array}{cc}
	\widetilde a_k &\widetilde  b_k \\\widetilde c_k & -\widetilde a_k \end{array} \right) 
	\lambda^{-k} + \left( \begin{array}{cc}
	\eta & 0 \\0 & \eta \end{array} \right)$$  
$$  = \sum_{k=0}^{n+1} \left( \begin{array}{cc}
	a_k & b_k \\ c_k & -a_k \end{array} \right) 
	\lambda^{n+1-k} + \left( \begin{array}{cc}
	\eta & 0 \\ 0 & \eta \end{array} \right)  
	 +\alpha \sum_{j=1}^N \frac{1}{\lambda-\lambda_j}
	\left( \begin{array}{cc}
	\phi_j\phi_{j,x} & -{\phi_j}^2 \\ 
	{\phi^2_{j,x}} & -\phi_j\phi_{j,x} 
	\end{array} \right),$$
where $\eta$ is a constant, also satisfies the adjoint representation 
(\ref{eqn:p1}), i.e.
\begin{equation}
\label{eqn:p9}
N^{(n)}_x=[U, N^{(n)}],
\end{equation}
which gives rise to the Lax representation of the constrained flow (\ref{eqn:p8}).

The KdV hierarchy with self-consistent 
sources is given by\cite{Zeng94, Zeng96}
\begin{subequations}
\label{eqn:p10}
\begin{equation}
\label{eqn:p10a}
	u_{t_n}=D\left[ \frac{\delta H_n}{\delta u} -
	2\alpha\sum_{j=1}^N 
	\frac{\delta \lambda_j}{\delta u} \right]\equiv
	D\left[ -2b_{n+2}-2\alpha\sum_{j=1}^N 
	{\phi_j}^2 \right],
\end{equation}
\begin{equation}
\label{eqn:p10b}
	\phi_{j,xx}+(\lambda_j+u)\phi_j=0,\qquad j=1,\cdots,N.
\end{equation}
\end{subequations}
Since the high-order constrained flows (\ref{eqn:p8}) are just 
the stationary equations of the KdV hierarchy with self-consistent sources (\ref{eqn:p10}),
it is obvious that the zero-curvature representation for the 
KdV hierarchy with self-consistent
sources (\ref{eqn:p10}) is given by 
\begin{equation}
\label{ss1}
	U_{t_n}-N^{(n)}_x+ [U, N^{(n)}]=0,
\end{equation}
with the auxiliary linear problems 
\begin{equation}
\label{eqn:KdVH-Source-Zero}
	{\left( \begin{array}{c}
	\psi \\ \psi_x \end{array} \right)}_x
	= U {\left( \begin{array}{c}
	\psi \\ \psi_x \end{array} \right)},\qquad 
{\left( \begin{array}{c}
	\psi \\ \psi_x \end{array} \right)}_{t_n}
	= N^{(n)} {\left( \begin{array}{c}
	\psi \\ \psi_x \end{array} \right)},  
\end{equation}
or equivalently
\begin{subequations}
\label{a5}
\begin{equation}
\label{a5a}
	\psi_{xx}+(\lambda+u) \psi =0,
\end{equation}
\begin{equation}
\label{a5b}
	\psi_{t_n}=A^{(n)}\psi+
	\eta\psi +\alpha \sum_{j=1}^N \frac{1}{\lambda-\lambda_j}
	\phi_j(\phi_{j,x}\psi-\phi_j\psi_x).
\end{equation}
\end{subequations}
Let assume that all products $\phi_j\psi$ decay at $x=-\infty$ and that $\p^{-1}
=\int_{-\infty}^x\cdot dx.$ It is easy to find from (\ref{eqn:p10b}) and 
(\ref{a5a}) that 
\begin{equation}
\label{a6}
	 \frac{1}{\lambda-\lambda_j}(\phi_{j,x}\psi-\phi_j\psi_x)
=\p^{-1}\phi_j\psi.
\end{equation}
Let denote
$$B_N=\alpha \sum_{j=1}^N \phi_j\p^{-1}\phi_j.$$
Then equation (\ref{a5b}) can be rewritten as 
\begin{equation}
\label{a5c}
	\psi_{t_n}=Q^{(n, N)}\psi\equiv A^{(n)}\psi+
	\eta\psi + B_N\psi.
\end{equation}

When $n=1,$ the equation (\ref{eqn:p10})
gives the KdV equation with self-consistent sources
\begin{subequations}
\label{a9}
\begin{equation}
\label{a9a}
	u_{t_1}=-\frac{1}{4}(6uu_x+u_{xxx})-
	2\alpha D\sum_{j=1}^N {\phi_j}^2,
\end{equation}
\begin{equation}
\label{a9b}
	\phi_{j,xx}+(\lambda_j+u)\phi_j=0,\qquad j=1,\cdots,N,
\end{equation}
\end{subequations}
and the auxiliary linear problem reads
\begin{subequations}
\label{a10}
\begin{equation}
\label{a10a}
\psi_{xx}+(\lambda+u) \psi =0,
\end{equation}
\begin{equation}
\label{a10b}
\psi_{t_1}=(\frac{1}{4} u_x +
	\eta)\psi +(\lambda-\frac{1}{2}u)\psi_x+\alpha \sum_{j=1}^N B_j\psi.
\end{equation}
\end{subequations}
  
\section{The Darboux transformation for the KdV  hierarchy}
\setcounter{equation}{0}
\hskip\parindent
In this section we recall the Darboux transformation for the KdV  hierarchy
(see \cite {salle} for a review, \cite{zakharov,orlov,chen}).

(1) The Darboux transformation for the KdV hierarchy.

Assume that $u$ be the solution of the n-th KdV equation (\ref{eqn:p6}) and 
denote the fixed solution of (\ref{a1}) and (\ref{a2}) with $\la=\xi$ by 
$f=f(x,t,\xi)$. The Darboux transformation (DT) is defined by
\begin{subequations}
\label{q1}
\begin{equation}
\label{q1a}
	\tilde \phi=\phi_x-\frac{f_x}f\phi,
\end{equation}
\begin{equation}
\label{q1b}
\tilde u=u+2\p^2\text{ln} f.
\end{equation}
 \end{subequations}
It is known that the 
 Schr\"{o}dinger equation (\ref{a1}) and (\ref{a2}) are covariant with respect 
to the action of the Darboux transformation ({\ref{q1}), 
namely $\tilde \phi, \tilde u$ satisfy
\begin{equation}
\label{q2}
	\tilde\phi_{xx}+(\lambda+\tilde u)\tilde \phi =0,
\end{equation}
\begin{equation}
\label{q3}
\tilde\phi_{t_n}=\tilde A^{(n)}\tilde\phi\equiv A^{(n)}(\tilde u, \la)\tilde \phi,
	\end{equation}
and $\tilde u$ satisfies the n-th KdV equation (\ref{eqn:p6}). 
Eqs. (\ref{a2}), (\ref{q1}) and (\ref{q3}) imply that
\begin{equation}
\label{q33}
\tilde\phi_{t_n}=[\phi_x-\frac{f_x}f\phi]_{t_n}=(A^{(n)}\phi)_x
-(\frac {A^{(n)}f}{f})_x\phi-\frac{f_x}fA^{(n)}\phi=\tilde A^{(n)}\tilde\phi.
\end{equation}
So the covariance of (\ref{a1}) and (\ref {a2}) with respect to 
the action of DT  (\ref{q1}) leads to the following lemma.

\begin{lemma} If $u$ is the solution of the n-th KdV equation (\ref{eqn:p6} 
and $f$ is a solution of (\ref{a1}) and (\ref{a2}) with $\la=\xi$ 
and   the Darboux transformation is given by (\ref{q1}), then the formula (\ref{q33}) holds.
\end{lemma}

We now construct the binary Darboux transformation.

(2) The first binary Darboux transformation.

Also, it is known
that the linearly independent solution of (\ref{a1}) and (\ref{a2}) 
with $\la=\xi$ is given by the Liouville formula
\begin{equation}
\label{q5}
	g=f\p^{-1}\frac 1{f^2}.
\end{equation}
The DT (\ref{q1}) implies that
\begin{equation}
\label{q6}
\tilde g=g_x-\frac{f_x}f g=\frac 1 f
\end{equation}
is one of solutions of (\ref{q2}) and (\ref{q3}) with $\la=\xi$ and 
$\tilde u$ given by (\ref{q1b}). The linearly independent solution 
$\tilde g_1$ of (\ref{q2}) and (\ref{q3}) with $\la=\xi$ is 
once more given by the Liouville formula
\begin{equation}
\label{q7}
	\tilde g_1=\tilde g\p^{-1}\frac 1{\tilde g^2}=\frac 1f\p^{-1} f^2.
\end{equation}

 By using $f$ and $\tilde g_1$, performing two-times repeated 
DT of (\ref{q1}) (notice that the right side of (\ref{q1a}) 
can be added a constant factor) and using  (\ref{a6}) give rise to
the binary Darboux transformation
\begin{subequations}
\label{q8}
\begin{eqnarray}
\label{q8a}
	&\bar \phi=\frac 1{\xi-\la}[\tilde \phi_x
-\frac{\tilde g_{1x}}{\tilde g_1}\tilde\phi]\nonumber\\
&=\frac 1{\xi-\la}[\xi\phi-\la\phi+\frac f{\p^{-1} f^2}(f_x\phi-f\phi_x)]
=\phi-\frac f{\p^{-1}f^2}\p^{-1}(f\phi),
\end{eqnarray}
\begin{equation}
\label{q8b}
\bar u=\tilde u+2\p^2 \text{ln}  \tilde g_1=u+2\p^2\text{ln} (\p^{-1}f^2).
\end{equation}
 \end{subequations}
Obviously, the equation (\ref{a1}) and (\ref{a2}) are covariant 
with respect to the action of the binary DT (\ref{q8}), namely $\bar \phi, \bar u$ satisfy 
\begin{equation}
\label{q9}
	\bar\phi_{xx}+(\lambda+\bar u)\bar \phi =0,
\end{equation}
\begin{equation}
\label{q10}
	\bar\phi_{t_n}=\bar A^{(n)}\bar\phi\equiv A^{(n)}(\bar u, \la)\bar \phi,
\end{equation}
and this $\bar u$ satisfies the n-th KdV equation (\ref{eqn:p6}).
It is found from (\ref{q8}), (\ref {a2}) and (\ref{q10}) that
\begin{eqnarray}
\label{q12}
&\bar\phi_{t_n}=\bar A^{(n)}\bar\phi=[\phi-\frac f{\p^{-1}f^2}\p^{-1}(f\phi)]_{t_n}\nonumber\\
&=A^{(n)}\phi-\frac 1{\p^{-1}f^2}\{[A^{(n)}f-\frac {2f}{\p^{-1}f^2}\p^{-1}(fA^{(n)}f)]
\p^{-1}(f\phi)
+f\p^{-1}[fA^{(n)}\phi+\phi A^{(n)}f]\}.
\end{eqnarray}
So the covariance of (\ref{a1}) and (\ref {a2}) with respect to the action of 
binary DT  (\ref{q8}) leads to the following lemma.

\begin{lemma} If $u$ is the solution of the n-th KdV equation (\ref{eqn:p6} 
and $f$ is a solution of (\ref{a1}) and (\ref{a2}) with $\la=\xi$ 
and   the binary DT is given by (\ref{q8}), then the formula (\ref{q12}) holds.
\end{lemma}

(3) The second binary Darboux transformation.

 Also the combination of $\tilde g$ and $\tilde g_1$ gives a  solution of 
(\ref{q2}) and (\ref{q3}) with $\la=\xi$ 
\begin{equation}
\tilde g_2=\tilde g+\tilde g_1 =\frac 1f(1+\p^{-1}f^2).
\end{equation}
By using $f$ and $\tilde g_2$, performing two-times repeated DT (\ref{q1}) 
leads to second binary Darboux transformation 
\begin{subequations}
\label{q13}
\begin{eqnarray}
\label{q13a}
	&\bar \phi=\frac 1{\xi-\la}[\tilde \phi_x-\frac{\tilde g_{2x}}{\tilde g_2}
\tilde\phi]\nonumber\\
&=\frac 1{\xi-\la}[\xi\phi-\la\phi+\frac f{1+\p^{-1} f^2}(f_x\phi-f\phi_x)]
=\phi-\frac f{1+\p^{-1}f^2}\p^{-1}(f\phi),
\end{eqnarray}
\begin{equation}
\label{q13b}
\bar u=\tilde u+2\p^2 \text{ln} \tilde g_2=u+2\p^2\text{ln} (1+\p^{-1}f^2).
\end{equation}
 \end{subequations}
Also the equation (\ref{a1}) and (\ref{a2}) are covariant with respect 
to the action of the binary DT (\ref{q13}), 
namely $\bar \phi, \bar u$ satisfy (\ref{q9}) and (\ref{q10}), $\bar u$ 
satisfies the n-th KdV equation (\ref{eqn:p6}).  
Similarly, the covariance of (\ref{a1}) and (\ref {a2}) 
with respect to the action of binary DT  (\ref{q13}) leads to the following lemma.

\begin {lemma}
 If $u$ is a solution  of the n-th KdV equation (\ref{eqn:p6} 
and $f$ is a solution of (\ref{a1}) and (\ref{a2}) with $\la=\xi$ and   
the binary DT is given by (\ref{q13}), then the following formula holds
\end{lemma}
\begin{eqnarray}
\label{q14}
&\bar\phi_{t_n}=\bar A^{(n)}\bar\phi=[\phi-\frac f{1+\p^{-1}f^2}\p^{-1}(f\phi)]_{t_n}\nonumber\\
&=A^{(n)}\phi-\frac 1{1+\p^{-1}f^2}\{[A^{(n)}f-\frac {2f}{1+\p^{-1}f^2}
\p^{-1}(fA^{(n)}f)]\p^{-1}(f\phi)
+f\p^{-1}[fA^{(n)}\phi+\phi A^{(n)}f]\}.\nonumber\\
&\ \
\end{eqnarray}

\section{The  Darboux transformation for the KdV  hierarchy with self-consistent sources}
\setcounter{equation}{0}
\hskip\parindent
Based on the Darboux transformation for the KdV hierarchy, 
we now construct Darboux transformation and two binary Darboux transformations 
for the KdV  hierarchy with self-consistent sources. 
The first binary Darboux transformation is an auto-B\"{a}cklund 
transformation for the n-th KdV  equation with self-consistent 
sources (\ref{eqn:p10}). The second one is a B\"{a}cklund transformation 
relating two n-th KdV  equations with self-consistent sources (\ref{eqn:p10}) 
with degree $N$ and $N+1$, respectively.

(1) Darboux transformation for the KdV hierarchy with sources.
\begin{theorem}
 Assume that $u, \phi_1,...,\phi_N$ be the solution of 
the n-th KdV equation with self-consistent sources  (\ref{eqn:p10}) 
and $f_1$ satisfies (\ref{a5a}) and (\ref{a5c}) with $\la=\xi_1$, 
then the Darboux  transformation is defined by
\begin{subequations}
\label{e1}
\begin{equation}
\label{e1a}
\bar \psi=\psi_x-\frac {f_{1,x}}{f_1}\psi,
\end{equation}
\begin{equation}
\label{e1b}
\bar u=u+2\p^2\text{ln} f_1,
\end{equation}
\begin{equation}
\label{e1c}
	\bar \phi_j=\frac 1{\sqrt{\la_j-\xi_1}}[\phi_{j,x}-\frac {f_{1,x}}{f_1}\phi_j],
\qquad j=1,...,N,
\end{equation}
\end{subequations}
the Lax representation (\ref{a5a}) and (\ref{a5c}) are covariant with respect to 
the  Darboux transformation (\ref{e1}). Namely $\bar u, \bar\psi, \bar\phi_j, j=1,...,N,$ 
satisfy
\begin{equation}
\label{c2}
	\bar\psi_{xx}+(\lambda+\bar u)\bar \psi =0,
\end{equation}
\begin{equation}
\label{c3}
	\bar\psi_{t_n}=\bar Q^{(n, N)}\bar \psi=\bar A^{(n)}\bar\psi+\eta\bar\psi 
+\bar B_N\bar\psi
\equiv A^{(n)}(\bar u, \la)\bar \psi+\eta\bar\psi +\alpha \sum_{j=1}^N\bar\phi_j
\p^{-1}(\bar\phi_j \bar\psi),
\end{equation}
and $\bar u, \bar\phi_1,\cdots,\bar\phi_N$ satisfy the n-th KdV equation 
with self-consistent sources (\ref{eqn:p10})
\begin{subequations}
\label{c4}
\begin{equation}
\label{c4a}
	\bar u_{t_n}=D\left[ -2\bar b_{n+2}-2\alpha\sum_{j=1}^N 
	\bar\phi_j^2 \right],
\end{equation}
\begin{equation}
\label{c4b}
	\bar\phi_{j,xx}+(\lambda_j+\bar u)\bar\phi_j=0,\qquad j=1,\cdots,N.
\end{equation}
\end{subequations}
\end{theorem}

{\bf{Proof}} Based on the results in the previous section, it is obvious that (\ref{c2}) 
and (\ref{c4b}) hold. In order to prove the (\ref{c3}) we need to show the following equality
\begin{equation}
\label{c44}
\tilde\psi_{t_n}=[\psi_x-\frac{f_{1,x}}{f_1}\psi]_{t_n}=(Q^{(n, N)}\psi)_x
-(\frac {Q^{(n, N)}f_1}{f_1})_x\psi-\frac{f_{1,x}}{f_1}Q^{(n, N)}\psi
=\tilde Q^{(n, N)}\tilde\psi.
\end{equation}
 The Lemma 3.1 implies that  equality (\ref{q33}) with $\phi$ replaced by 
$\psi$ holds. So we only need to check the terms containing $\phi_1,...,\phi_N$ 
in the equality (\ref{c44}), i.e., to show the following equality 
\begin{eqnarray}
\label{c5}
&(B_N\psi)_x-(\frac {B_Nf_1}{f_1})_x\psi-\frac {f_{1,x}}{f_1}B_N\psi=\bar B_N\bar\psi.
\end{eqnarray}
Using (\ref{e1}) and (\ref{a6}), we have
\begin{eqnarray}
\label{c7}
&\bar B_N\bar \psi=\alpha\sum_{j=1}^N\bar \phi_j\p^{-1}[\frac 1{\sqrt{\la_j-\xi_1}}
(\phi_{j,x}-\frac {f_{1,x}}{f_1}\phi_j)(\psi_x-\frac {f_{1,x}}{f_1}\psi)]\nonumber\\
&=\alpha\sum_{j=1}^N\frac 1{\sqrt{\la_j-\xi_1}}\bar \phi_j[\phi_{j,x}\psi-\p^{-1}
(\phi_{j,xx}\psi)-\frac {f_{1,x}}{f_1}\phi_j\psi+\p^{-1}(\frac {f_{1,xx}}{f_1}\phi_j\psi)]
\nonumber\\
&=\alpha\sum_{j=1}^N\frac 1{\sqrt{\la_j-\xi_1}}\bar \phi_j[
\phi_{j,x}\psi-\frac {f_{1,x}}{f_1}\phi_j\psi+(\la_j-\xi_1)\p^{-1}(\phi_j\psi)],
\end{eqnarray}
and
\begin{eqnarray}
\label{c8}
&\text{the left terms in (\ref{c5})}=
\alpha\sum_{j=1}^N[\phi_{j,x}\p^{-1}(\phi_j\psi)-\frac 1{f_1}\psi\phi_{j,x}
\p^{-1}(f_1\phi_j)+\frac {f_{1,x}}{f_1^2}\psi\phi_j\p^{-1}(f_1\phi_j)\nonumber\\
&-\frac{f_{1,x}}{f_1}\phi_j\p^{-1}(\phi_j\psi)]
=\alpha\sum_{j=1}^N\sqrt{\la_j-\xi_1}\bar\phi_{j}[\p^{-1}(\phi_j\psi)
-\frac 1{f_1}\psi\p^{-1}(f_1\phi_j)]\nonumber\\
&=\alpha\sum_{j=1}^N\sqrt{\la_j-\xi_1}\bar\phi_{j}[\p^{-1}(\phi_j\psi)+\frac 1{f_1(\la_j-\xi_1)}
\psi(f_1\phi_{j,x}-f_{1,x}\phi_j)].
\end{eqnarray}
Comparing (\ref{c7}) with (\ref{c8}), it is immediately found that equality (\ref{c5}) holds.
 The equation (\ref{c2}) and (\ref{c3}) lead to (\ref{c4a}).
This completes the proof.

(2) The first binary Darboux transformation for the KdV hierarchy with sources.
\begin{theorem}
 Assume that $u, \phi_1,...,\phi_N$ be the solution of the n-th KdV equation 
with self-consistent sources  (\ref{eqn:p10}) and $f_1$ satisfies (\ref{a5a}) and (\ref{a5c}) 
with $\la=\xi_1$, then the first binary Darboux  transformation is defined by
\begin{subequations}
\label{e2}
\begin{equation}
\label{e2a}
\bar \psi=\psi-\frac {f_1}{\p^{-1}f_1^2}\p^{-1}(f_1\psi),
\end{equation}
\begin{equation}
\label{e2b}
\bar u=u+2\p^2\text{ln} (\p^{-1}f_1^2),
\end{equation}
\begin{equation}
\label{e2c}
	\bar \phi_j=\phi_j-\frac {f_1}{\p^{-1}f_1^2}\p^{-1}(f_1\phi_j),\qquad j=1,...,N,
\end{equation}
\end{subequations}
the Lax representation (\ref{a5a}) and (\ref{a5c}) are covariant with respect to 
the binary Darboux transformation (\ref{e2}). Namely $\bar u, \bar\psi, \bar\phi_j, j=1,...,N,$ satisfy
(\ref{c2}), (\ref{c3}) 
and  the n-th KdV equation with self-consistent sources (\ref{c4}).
\end{theorem}

{\bf{Proof}} It is obvious that (\ref{c2}) and (\ref{c4b}) hold. Similarly, in order 
to prove the (\ref{c3}) we need to show the equality (\ref{q12}) with $\phi, A^{(n)}$ 
replaced by $\psi, Q^{(n, N)}$. In fact, using Lemma 3.2, we only need to check the terms 
containing $\phi_1,...,\phi_N$ in the equality, i.e., to show the following equality 
\begin{eqnarray}
\label{e5}
&B_N\psi-\frac 1{\p^{-1}f_1^2}[B_Nf_1-2\frac 1{\p^{-1}f_1^2}f_1\p^{-1}(f_1B_Nf_1)]
\p^{-1}(f_1\psi)\nonumber\\
&-\frac {f_1}{\p^{-1}f_1^2}\p^{-1}[f_1B_N\psi
+\psi B_Nf_1]=\bar B_N\bar\psi.
\end{eqnarray}
Notice that
\begin{eqnarray}
\label{e6}
&\p^{-1}[\frac {f_1^2}{(\p^{-1}f_1^2)^2}(\p^{-1}(f_1\phi_j))(\p^{-1}(f_1\psi))]=
-\frac 1{\p^{-1}f_1^2}(\p^{-1}(f_1\phi_j))(\p^{-1}(f_1\psi))\nonumber\\
&+\p^{-1}[\frac {f_1\phi_j}{\p^{-1}f_1^2}\p^{-1}(f_1\psi)+
\frac {f_1\psi}{\p^{-1}f_1^2}\p^{-1}(f_1\phi_j)].
\end{eqnarray}
Using (\ref{e2}) and (\ref{e6}), we have
\begin{eqnarray}
\label{e7}
&\bar B_N\bar \psi=\alpha\sum_{j=1}^N\bar \phi_j\p^{-1}(\bar\phi_j\bar\psi)\nonumber\\
&=\alpha\sum_{j=1}^N\bar \phi_j\p^{-1}[(\phi_j-\frac {f_1}{\p^{-1}f_1^2}\p^{-1}(f_1\phi_j))
(\psi-\frac {f_1}{\p^{-1}f_1^2}\p^{-1}(f_1\psi))]\nonumber\\
&=\alpha\sum_{j=1}^N\bar \phi_j[\p^{-1}(\phi_j\psi)-\frac 1{\p^{-1}f_1^2}(\p^{-1}(f_1\phi_j))
(\p^{-1}(f_1\psi))],
\end{eqnarray}
and
\begin{eqnarray}
\label{e8}
&\text{the left terms in (\ref{e5})}=
\alpha\sum_{j=1}^N\{\phi_j\p^{-1}(\phi_j\psi)-\frac 1{\p^{-1}f_1^2}[\phi_j(\p^{-1}(f_1\phi_j))
(\p^{-1}(f_1\psi))\nonumber\\
&-2\frac {f_1}{\p^{-1}f_1^2}\p^{-1}(f_1\phi_j\p^{-1}(f_1\phi_j))(\p^{-1}(f_1\psi))
+f_1\p^{-1}(f_1\phi_j\p^{-1}(\phi_j\psi))
+f_1\p^{-1}(\phi_j\psi\p^{-1}(\phi_j f_1))]\}\nonumber\\
&=\alpha\sum_{j=1}^N[\phi_j\p^{-1}(\phi_j\psi)-\frac {\phi_j}{\p^{-1}f_1^2}
(\p^{-1}(f_1\phi_j))(\p^{-1}(f_1\psi))\nonumber\\
&+\frac {f_1}{(\p^{-1}f_1^2)^2}(\p^{-1}(f_1\phi_j))^2(\p^{-1}(f_1\psi))
-\frac {f_1}{\p^{-1}f_1^2}(\p^{-1}(f_1\phi_j))(\p^{-1}(\phi_j\psi))].
\end{eqnarray}
By substituting (\ref{e2c}) into (\ref{e7}) and comparing it with (\ref{e8}),
 it is immediately found that equality (\ref{e5}) holds. The equation (\ref{c2}) 
and (\ref{c3}) lead to (\ref{c4a}).
This completes the proof.

(3) The second binary Darboux transformation for the KdV hierarchy with sources.
\begin{theorem}
 Assume that $u, \phi_1,...,\phi_N$ be the solution of the n-th KdV equation 
with self-consistent sources  (\ref{eqn:p10}),  $f_1\equiv \phi_{N+1}$ satisfies (\ref{a5a}) 
and (\ref{a5c}) with $\la=\la_{N+1}$ and $\eta=-\frac 12\alpha$, 
then the  second binary Darboux transformation is defined by 
\begin{subequations}
\label{c11}
\begin{equation}
\label{c11a}
\bar \psi=\psi-\frac {f_1}{1+\p^{-1}f_1^2}\p^{-1}(f_1\psi)=\psi-\bar\phi_{N+1}\p^{-1}(f_1\psi),
\end{equation}
\begin{equation}
\label{c11b}
\bar u=u+2\p^2\text{ln} (1+\p^{-1}f_1^2),
\end{equation}
\begin{equation}
\label{c11c}
	\bar \phi_j=\phi_j-\frac {f_1}{1+\p^{-1}f_1^2}\p^{-1}(f_1\phi_j)
=\phi_j-\bar \phi_{N+1}\p^{-1}(f_1\phi_j),\qquad j=1,...,N,
\end{equation}
where
\begin{equation}
\label{c11d}
	\bar \phi_{N+1}=\frac {f_1}{1+\p^{-1}f_1^2}, \qquad f_1=\phi_{N+1},
\end{equation}
\end{subequations}
and the binary Darboux transformation (\ref{c11}) transforms
 the Lax representation (\ref{a5a}) and (\ref{a5c}) with $\eta=-\frac 12\alpha$ 
into the following Lax representation
\begin{equation}
\label{c12}
	\bar\psi_{xx}+(\lambda+\bar u)\bar \psi =0,
\end{equation}
\begin{equation}
\label{c13}
	\bar\psi_{t_n}=\bar Q^{(n, N+1)}\bar \psi=\bar A^{(n)}\bar\psi
-\frac 12\alpha\bar\psi +\bar B_{N+1}\bar\psi
\equiv A^{(n)}(\bar u, \la)\bar \psi-\frac 12\alpha\bar\psi 
+\alpha \sum_{j=1}^{N+1}\bar\phi_j\p^{-1}(\bar\phi_j \bar\psi),
\end{equation}
and $\bar u, \bar\phi_1,\cdots,\bar\phi_{N+1}$ satisfies the n-th KdV equation 
with self-consistent sources (\ref{eqn:p10}) with degree $N+1$
\begin{subequations}
\label{c14}
\begin{equation}
\label{c14a}
	\bar u_{t_n}=D\left[ -2\bar b_{n+2}-2\alpha\sum_{j=1}^{N+1} 
	{\bar\phi_j}^2 \right],
\end{equation}
\begin{equation}
\label{c14b}
	\bar\phi_{j,xx}+(\lambda_j+\bar u)\bar\phi_j=0,\qquad j=1,\cdots,N+1.
\end{equation}
\end{subequations}
\end{theorem}
{\bf{Proof}}  It is easy to see that (\ref{c11c}) holds for $j=N+1$.
 So, based on the results in previous section, it is obvious that (\ref{c12}) 
and (\ref{c14b}) hold. Similarly, in order to prove (\ref{c13}), 
by using Lemma 3.3 one only needs to check the terms containing 
$\phi_1,...,\phi_N,\bar\phi_{N+1}$ in the equality  (\ref {q14}) 
with $\phi, A^{(n)}$ replaced by $\psi, Q^{(n, N)},$  i.e., to show the following equality 
\begin{eqnarray}
\label{c16}
&B_N\psi-\frac 1{1+\p^{-1}f_1^2}[B_Nf_1-\frac 12\alpha\bar\phi_{N+1}(1-\p^{-1}f_1^2)
-2 \bar\phi_{N+1}\p^{-1}(f_1B_Nf_1)]\p^{-1}(f_1\psi)\nonumber\\
&-\bar\phi_{N+1}\p^{-1}[\psi B_N f_1-\frac 12\alpha f_1\psi+f_1B_N\psi]
=\bar B_{N+1}\bar\psi.
\end{eqnarray}
Notice that
\begin{equation}
\p^{-1}(\bar\phi_{N+1}\bar\psi)=\p^{-1}[\frac {f_1}{1+\p^{-1}f_1^2}\psi
-(\frac {f_1}{1+\p^{-1}f_1^2})^2\p^{-1}(f_1\psi)]=\frac 1{1+\p^{-1}f_1^2}\p^{-1}(f_1\psi).
\end{equation}
By means of (\ref{e6}) and (\ref{e7}) with $\p^{-1}f_1^2$ replaced by $(1+\p^{-1}f_1^2)$ one gets
\begin{eqnarray}
\label{c17}
&\bar B_{N+1}\bar \psi=\alpha\sum_{j=1}^{N+1}\bar \phi_j\p^{-1}(\bar\phi_j\bar\psi)\nonumber\\
&=\alpha\sum_{j=1}^N\bar \phi_j[\p^{-1}(\phi_j\psi)-
\frac 1{1+\p^{-1}f_1^2}(\p^{-1}(f_1\phi_j))\p^{-1}(f_1\psi)]+
\frac {\alpha\bar\phi_{N+1}}{1+\p^{-1}f_1^2}\p^{-1}(f_1\psi).
\end{eqnarray}
Using (\ref{e8}) with  $\p^{-1}f_1^2$ replaced by $(1+\p^{-1}f_1^2)$ it is found 
\begin{eqnarray}
\label{c18}
&\text{the left terms in (\ref{c16})}
=\alpha\sum_{j=1}^N\{\phi_j\p^{-1}(\phi_j\psi)-
\frac 1{1+\p^{-1}f_1^2}[{\phi_j}(\p^{-1}(f_1\phi_j))\p^{-1}(f_1\psi)\nonumber\\
&-\bar\phi_{N+1}(\p^{-1}(f_1\phi_j))^2\p^{-1}(f_1\psi)]
-\bar\phi_{N+1}(\p^{-1}(f_1\phi_j))\p^{-1}(\phi_j\psi)\}
+\frac {\alpha\bar\phi_{N+1}}{1+\p^{-1}f_1^2}\p^{-1}(f_1\psi).
\end{eqnarray}
By substituting (\ref{c11c}) into (\ref{c17}) and comparing it with (\ref{c18}),
 it is easy to see that equality (\ref{c16}) holds. The equations (\ref{c12})
 and (\ref{c13}) yield (\ref{c14}).
This completes the proof.

{\bf{Remark}} The binary Darboux transformation defined by (\ref{c11}) is
 a non auto-B\"{a}cklund transformation relating the two n-th KdV equations 
with self-consistent sources (\ref{eqn:p10}) and (\ref{c14}). 
This Darboux transformation can be used to construct the soliton solution for (\ref{eqn:p10}).

For example, in order to find one soliton solution for the KdV equation
 with self-consistent sources (\ref{a9}) with $N=1$, we start from
 the solution $u=0 $ for the KdV equation with self-consistent sources (\ref{a9})
 with $N=0$. The solution for (\ref{a10}) with $N=0, u=0, \la=-k^2, k>0, \eta=-\frac 12\alpha$ reads
$$\phi_1=ce^{kx-k^3t-\frac 12\alpha t}.$$
Then one finds from (\ref{c11}) that
$$\bar u=2k^2\text{sech}^2(kx-k^3t-\frac 12\alpha t+x_0),$$
$$\phi_1=\frac 12\sqrt{2k}\text{sech}(kx-k^3t-\frac 12\alpha t+x_0),$$
which is the one soliton solution for the KdV equation with self-consistent sources (\ref{a9}) with $N=1$.

\section{The m-times repeated binary Darboux transformation for 
the KdV  hierarchy with self-consistent sources}
\setcounter{equation}{0}
\hskip\parindent
It is evident that the Darboux transformation can be applied to (\ref{c2}), 
(\ref{c3}) and (\ref{c12}), (\ref{c13}) once more to produce some new solutions 
for the KdV  hierarchy with self-consistent sources. 

(1) The m-times repeated  second binary Darboux transformation.

 Assume that
$f_1,...,f_m$ be solution of (\ref{a5a}) and (\ref{a5c}) 
with $\la=\la_{N+1},...,\la_{N+m},$ respectively. We use $u[i], \psi[i],f_j[i], \phi_j[i]$ 
to denote the action of i-times repeated binary Darboux transformation of (\ref{c11}) 
on the initial solution $u, \psi, f_j, \phi_j.$ We have
\begin{equation}
\label{d1}
	f_j[i]_{xx}+(\lambda_j+ u[i])f_j[i] =0,
\end{equation}
\begin{equation}
\label{d2}
	f_{j,t_n}[i]= Q^{(n, N+i)}[i]f_j[i].
\end{equation}
We define two integral types of the Wronskian determinant of $k$ functions $g_1,...,g_k$ 
in a similar way as in \cite{chen} by
$$W_1(g_1,...,g_k)=det F,\qquad \qquad W_2(g_1,...,g_k)=det G,$$
where
$$F_{ij}=\delta_{ij}+\p^{-1}(g_ig_j),\qquad\quad i,j=1,...,k,$$
$$G_{ij}=\delta_{ij}+\p^{-1}(g_ig_j), \quad i=1,...,k-1,\quad j=1,...,k,
\quad\qquad G_{kj}=g_j,\quad j=1,...,k. $$
\begin{lemma}
For arbitrary integers $l, k (1\leq l\leq m-1, 1\leq k\leq m-l),$ we have
\begin{equation}
\label{d3}
W_1(f_{l+1}[l],...,f_{l+k}[l])=\frac {W_1(f_{l}[l-1],f_{l+1}[l-1],...,
f_{l+k}[l-1])}{1+\p^{-1}f_l^2[l-1]},
\end{equation}
\begin{equation}
\label{d4}
W_2(f_{l+1}[l],...,f_{l+k}[l],\psi[l])=\frac {W_1(f_{l}[l-1],f_{l+1}[l-1],...,
f_{l+k}[l-1],\psi[l-1])}{1+\p^{-1}f_l^2[l-1]}.
\end{equation}
\end{lemma}
{\bf{Proof}} According to (\ref{c11}), we have
\begin{equation}
\label{d5}
	f_{l+j}[l]=f_{l+j}[l-1]-\frac {f_{l}[l-1]}{1+\p^{-1}f^2_l[l-1]}\p^{-1}(f_l[l-1]f_{l+j}[l-1]),
\end{equation}
so using (\ref{e6})
\begin{eqnarray}
\label{d6}
&F_{ij}=\delta_{ij}+\p^{-1}(f_{l+i}[l]f_{l+j}[l])=\delta_{ij}+\p^{-1}(f_{l+i}[l-1]f_{l+j}[l-1])
\nonumber\\
&-\frac {1}{1+\p^{-1}f^2_l[l-1]}(\p^{-1}f_{l+i}[l-1]f_{l}[l-1])(\p^{-1}f_l[l-1]f_{l+j}[l-1])
\equiv\delta_{ij}+a_{ij}-ba_{i0}a_{0j},
\end{eqnarray}
where
$$a_{ij}=\p^{-1}(f_{l+i}[l-1]f_{l+j}[l-1]), \qquad  b=\frac {1}{1+\p^{-1}f^2_l[l-1]}.$$
Then
$$W_1(f_{l+1}[l],...,f_{l+k}[l])=\text{det}(F_{ij})$$
$$= \left( \begin{array}{ccccc}
	1+a_{11}-ba_{10}a_{01} &a_{12}-ba_{10}a_{02}&a_{13}-ba_{10}a_{03}& \cdots &a_{1k}-ba_{10}a_{0k} \\
a_{21}-ba_{20}a_{01} &1+a_{22}-ba_{20}a_{02}&a_{23}-ba_{20}a_{03}& \cdots &a_{2k}-ba_{20}a_{0k} \\
\vdots&\vdots&\vdots&\ddots&\vdots\\
a_{k1}-ba_{k0}a_{01} &a_{k2}-ba_{k0}a_{02}&a_{k3}-ba_{k0}a_{03}& \cdots &1+a_{kk}-ba_{k0}a_{0k} \\
\end{array} \right)$$
$$ =\left( \begin{array}{cccc}
	1+a_{11} &a_{12}&\cdots &a_{1k} \\
a_{21}&1+a_{22}&\cdots&a_{2k} \\
\vdots&\vdots&\ddots&\vdots\\
a_{k1}&a_{k2}&\cdots &1+a_{kk} \\
\end{array} \right)
-ba_{01} \left( \begin{array}{ccccc}
	a_{10} &a_{12}& a_{13}&\cdots &a_{1k} \\
a_{20}&1+a_{22}&a_{23}&\cdots&a_{2k} \\
\vdots&\vdots&\vdots&\ddots&\vdots\\
a_{k0}&a_{k2}&a_{k3}& \cdots &1+a_{kk} \\
\end{array} \right)$$
$$-ba_{02} \left( \begin{array}{ccccc}
	1+a_{11} &a_{10}&a_{13}&\cdots &a_{1k} \\
a_{21}&a_{20}&a_{23}&\cdots&a_{2k} \\
\vdots&\vdots&\vdots&\ddots&\vdots\\
a_{k1}&a_{k0}&a_{k3}&\cdots &1+a_{kk} \\
\end{array} \right)-\cdots$$
$$-ba_{0k} \left( \begin{array}{ccccc}
	1+a_{11} &a_{12}&\cdots&a_{1(k-1)} &a_{10} \\
a_{21}&1+a_{22}&\cdots &a_{2(k-1)}&a_{20} \\
\vdots&\vdots&\ddots&\vdots&\vdots\\
a_{k1}&a_{k2}&\cdots &a_{k(k-1)}&a_{k0} \\
\end{array} \right)$$
$$=\frac {1}{1+a_{00}} \left( \begin{array}{ccccc}
	1+a_{00} &a_{01}&a_{02}&\cdots &a_{0k} \\
a_{10}&1+a_{11}&a_{12}&\cdots &a_{1k} \\
\vdots&\vdots&\vdots&\ddots&\vdots\\
a_{k0}&a_{k1}&a_{k2}&\cdots &1+a_{kk} \\
\end{array} \right)
=\frac {W_1(f_{l}[l-1],...,f_{l+k}[l-1])}{1+\p^{-1}f_l^2[l-1]}.$$
In the similar way the formula (\ref{d4}) can be proved. This completes the proof.

\begin{theorem}
Assume that $u, \phi_1,\cdots\phi_N$ is solution of 
 the n-th KdV equation with self-consistent sources (\ref{eqn:p10}), 
$f_1,...,f_m$ be solution of (\ref{a5a}) and (\ref{a5c}) with $\la=\la_{N+1},...,\la_{N+m}, $  
respectively, and $\eta=-\frac 12\alpha$.    
Then the m-times repeated binary Darboux transformation of (\ref{c11}) is given by
\begin{subequations}
\label{d6}
\begin{equation}
\label{d6a}
\psi[m]=\frac {W_2(f_1,..., f_m, \psi)}{W_1(f_1,..., f_m)},
\end{equation}
\begin{equation}
\label{d6b}
u[m]=u+2\p^2\text{ln} W_1(f_1,..., f_m),
\end{equation}
\begin{equation}
\label{d6c}
	\phi_j[m]=\frac {W_2(f_1,..., f_m, \phi_j)}{W_1(f_1,..., f_m)}, \qquad j=1,\cdots,N,
\end{equation}
\begin{equation}
\label{d6d}
	\phi_{N+j}[m]=\frac {W_2(f_1,..., f_m, f_j)}{W_1(f_1,..., f_m)}=
\frac {W_2(f_1,..., f_{j-1}, f_{j+1},..., f_m, f_j)}{W_1(f_1,..., f_m)}, \quad j=1,\cdots,m,
\end{equation}
\end{subequations}
and $u[m], \psi[m], \phi_1[m],\cdots,\phi_{N+m}[m]$ satisfy
\begin{equation}
\label{d7}
	\psi_{xx}[m]+(\lambda+ u[m])\psi[m] =0,
\end{equation}
\begin{equation}
\label{d8}
	\psi_{t_n}[m]= Q^{(n, N+m)}[m]\psi[m]=A^{(n)}(u[m], \la)\psi[m]-\frac 12\alpha\psi[m] 
+\alpha\sum_{j=1}^{N+m}\phi_j[m]\p^{-1}(\phi_j[m]\psi[m]),
\end{equation}
and
\begin{subequations}
\label{d9}
\begin{equation}
\label{d9a}
	 u_{t_n}[m]=D\left[ -2 b_{n+2}(u[m])-2\alpha\sum_{j=1}^{N+m} 
	\phi_j^2[m] \right],
\end{equation}
\begin{equation}
\label{d9b}
	\phi_{j,xx}[m]+(\lambda_j+ u[m])\phi_j[m]=0,\qquad j=1,\cdots,N+m.
\end{equation}
\end{subequations}
\end{theorem}
{\bf{proof}} Using  (\ref{c11}),   (\ref{d3}) and (\ref{d4}), one obtains
\begin{eqnarray}
&\psi[m]=\psi[m-1]-\frac {f_m[m-1]}{1+\p^{-1}f^2_m[m-1]}\p^{-1}(f_m[m-1]\psi[m-1])\nonumber\\
&=\frac {1}{1+\p^{-1}f^2_m[m-1]}W_2(f_m[m-1], \psi[m-1])
=\frac {W_2(f_m[m-1], \psi[m-1])}{W_1(f_m[m-1])}\nonumber\\
&=\frac {W_2(f_{m-1}[m-2], f_m[m-2], \psi[m-2])}{1+\p^{-1}f^2_{m-1}[m-2]}
\cdot\frac {1+\p^{-1}f^2_{m-1}[m-2]}{W_1(f_{m-1}[m-2],  f_{m}[m-2])}
=\cdots=\frac {W_2(f_1,..., f_m, \psi)}{W_1(f_1,..., f_m)},
\end{eqnarray}
\begin{eqnarray}
&u[m]=u[m-1]+2\p^2\text{ln} (1+\p^{-1}f^2_m[m-1])=u[m-1]+2\p^2\text{ln} W_1(f_m[m-1])\nonumber\\
&=u[m-2]+2\p^2\text{ln} (1+\p^{-1}f^2_{m-1}[m-2])+2\p^2\text{ln}
\frac {W_1(f_{m-1}[m-2], f_m[m-2])}{1+\p^{-1}f^2_{m-1}[m-2]}\nonumber\\
&=u[m-2]+2\p^2\text{ln}W_1(f_{m-1}[m-2], f_m[m-2])=\cdots=u+2\p^2\text{ln} W_1(f_1,..., f_m).
\end{eqnarray}
Similarly we can prove the (\ref{d6c}) and (\ref{d6d}). It is easy to find (\ref{d7}), (\ref{d8}) 
and (\ref{d9}) from the Proposition 4.3.

The m-times repeated binary Darboux transformation (\ref{d7}) provides a B\"{a}cklnud transformation 
relating two n-th KdV equations with self-consistent sources (\ref{eqn:p10}) with 
degree $N$ and $N+m$, respectively. We now use the N-times repeated 
binary Darboux transformation (\ref{d6}) to construct the N-soliton solution for 
the n-th KdV equation with self-consistent sources (\ref{eqn:p10}) with $\la_j=-k_j^2<0, k_j>0,
 j=1,\cdots,N.$ We start from (\ref{eqn:p10}) with $N=0$. Taking $N=0, u=0, \la=-k_j^2, 
\eta=-\frac 12\alpha$, then (\ref{a5a}) and (\ref{a5c}) reduce to
$$\psi_{xx}-k_j^2\psi=0,$$
$$\psi_{t_n}=(-1)^nk_j^{2n}\psi_x-\frac 12\alpha\psi,$$
which solution is given by
\begin{equation}
\label{d10}
f_j=e^{k_jx+(-1)^nk_j^{2n+1}t_n-\frac 12\alpha t_n+x_{0,j}}, \qquad j=1,\cdots,N.
\end{equation}
 Then according to the Proposition 5.1, the N-soliton solution for the n-th KdV 
equation with self-consistent sources (\ref{eqn:p10}) with $\la_j=-k_j^2<0, k_j>0, j=1,\cdots,N, $ 
is given by
\begin{equation}
\label{d11a}
u=2\p^2\text{ln} W_1(f_1,...,f_N),
\end{equation}
\begin{equation}
\label{d11b}
	\phi_{j}=\frac {W_2(f_1,...,f_N, f_j)}{W_1(f_1,...,f_N)}=
\frac {W_2(f_1,...,f_{j-1}, f_{j+1},...,f_N,f_j)}{W_1(f_1,...,f_N)}, \quad j=1,\cdots,N,
\end{equation}
where $f_j$ is given by (\ref{d10}).

(2) The m-times repeated first binary Darboux transformation.

 We define
\begin{equation}
\label{d12}
W_1(g_1,...,g_k)=det F,\qquad \qquad W_2(g_1,...,g_k)=det G,
\end{equation}
where
$$F_{ij}=\p^{-1}(g_ig_j),\qquad\quad i,j=1,...,k,$$
$$G_{ij}=\p^{-1}(g_ig_j), \quad i=1,...,k-1,\quad j=1,...,k,\quad\qquad G_{kj}=g_j,\quad j=1,...,k. $$

In the exactly same way we can prove the following theorem.
\begin{theorem}
Assume that $u, \phi_1,\cdots\phi_N$ is solution of 
 the n-th KdV equation with self-consistent sources (\ref{eqn:p10}), 
$f_1,...,f_m$ be the solutions of (\ref{a5a}) and (\ref{a5c}) with $\la=\xi_{1},...,\xi_{m},$ 
respectively. 
Then the m-times repeated binary Darboux transformation of (\ref{e2}) is given by 
\begin{subequations}
\label{d13}
\begin{equation}
\label{d13a}
\psi[m]=\frac {W_2(f_1,...,f_m, \psi)}{W_1(f_1,...,f_m)},
\end{equation}
\begin{equation}
\label{d13b}
u[m]=u+2\p^2\text{ln} W_1(f_1,...,f_m),
\end{equation}
\begin{equation}
\label{d13c}
	\phi_j[m]=\frac {W_2(f_1,...,f_m, \phi_j)}{W_1(f_1,...,f_m)}, \qquad j=1,\cdots,N,
\end{equation}
\end{subequations}
and $u[m], \psi[m], \phi_1[m],\cdots,\phi_{N}[m]$ satisfy
\begin{equation}
\label{d14}
	\psi_{xx}[m]+(\lambda+ u[m])\psi[m] =0,
\end{equation}
\begin{equation}
\label{d15}
	\psi_{t_n}[m]= Q^{(n, N)}[m]\psi[m],
\end{equation}
and
\begin{subequations}
\label{d16}
\begin{equation}
\label{d16a}
	 u_{t_n}[m]=D\left[ -2 b_{n+2}(u[m])-2\alpha\sum_{j=1}^{N} 
	\phi^2_j[m] \right],
\end{equation}
\begin{equation}
\label{d16b}
	\phi_{j,xx}[m]+(\lambda_j+ u[m])\phi_j[m]=0,\qquad j=1,\cdots,N.
\end{equation}
\end{subequations}
\end{theorem}

(3) The m-times repeated  Darboux transformation of (\ref{e1}).

 We define the Wronskian determinant $W$ by
\begin{equation}
\label{d17}
W_1(g_1,...,g_k)=det F,\qquad \qquad F_{ij}=\frac {\p^{i-1}g_j}{\p x^{i-1}},\qquad i,j=1,...,k.
\end{equation}
In the exactly same way we can prove the following theorem.
\begin{theorem}
Assume that $u, \phi_1,\cdots\phi_N$ is solution of 
 the n-th KdV equation with self-consistent sources (\ref{eqn:p10}), 
$f_1,...,f_m$ be the solutions of (\ref{a5a}) and (\ref{a5c}) with $\la=\xi_{1},...,\xi_{m},$ 
respectively. 
Then the m-times repeated  Darboux transformation of (\ref{e1}) is given by 
\begin{subequations}
\label{d18}
\begin{equation}
\label{d18a}
\psi[m]=\frac {W(f_1,...,f_m, \psi)}{W(f_1,...,f_m)},
\end{equation}
\begin{equation}
\label{d18b}
u[m]=u+2\p^2\text{ln} W(f_1,...,f_m),
\end{equation}
\begin{equation}
\label{d18c}
	\phi_j[m]=\frac {W(f_1,...,f_m, \phi_j)}{W(f_1,...,f_m)}, \qquad j=1,\cdots,N,
\end{equation}
\end{subequations}
and $u[m], \psi[m], \phi_1[m],\cdots,\phi_{N}[m]$ satisfy (\ref{d14}), (\ref{d15})
and (\ref{d16}).
\end{theorem}

\hskip\parindent

\section*{ Acknowledgment }
This work was in part supported by a grant from  the Research Grants Council of 
Hong Kong Special Administrative Region, China (Project no. 9040466) , and a grant from 
the City University of Hong Kong (Project no. 7001041), as well as by 
the Special Funds for Chinese Major Basic Research Project "Nonlinear Science" .

\hskip\parindent



\begin{thebibliography}{s99}
\bibitem{mel89a}
V. K. Mel'nikov, 
Commun. Math. Phys. 120, 451 (1989);

\bibitem{mel89b}
V. K. Mel'nikov, 
Commun. Math. Phys. 126, 201-215 (1989).

\bibitem{mel90a}
V. K. Mel'nikov, 
Inverse Problem 6, 233-246 (1990).


\bibitem{Kaup87}
D. J. Kaup, 
Phys. Rev. Lett. 59, 2063 (1987).


\bibitem{Lati90}
J. Leon  and A. Latifi, 
 J. Phys. A 23, 1385 (1990).

\bibitem{Clau91}
C. Claude, A. Latifi  and J. Leon,
 J. Math. Phys. 32, 3321 (1991).

\bibitem{Vlas91}
R. A. Vlasov and E. V. Doktorov E V,
Dokl. Akad. Nauk BSSR 26, 17 (1991).

\bibitem{Dokt83}
E. V. Doktorov and  R. A. Vlasov, 
Opt. Acta 30, 223 (1993).

\bibitem{Naka91}
M. Nakazawa, E. Yomada and H. Kubota, 
Phys. Rev. Lett. 66, 2625 (1991).

\bibitem{Dokt95} 
E. V. Doktorov and V. S. Shchesnovich, 
Phys. Lett. A  207, 153 (1995).

\bibitem{Shch96}
V. S. Shchesnovich and E. V. Doktorov,
 Phys. Lett. A 213, 23 (1996).

\bibitem{mel90}
V. K. Mel'nikov, 
J. Math. Phys. 31, 1106 (1990).

\bibitem{Leon89}
J. Leon, 
J. Math. Phys. 29, 2012 (1988);
 Phys. Lett. A 144, 444 (1990).

\bibitem{Anto92}
M. Antonowicz and S. R. Wojciechowski, 
Phys. Lett. A 165, 47 (1992).

\bibitem{Zeng93}
Y. B. Zeng, 
J. Phys. A: Math. Gen. 26, L273 (1993).

\bibitem{Zeng94}
Y. B. Zeng,
Physica D 73, 171 (1994).

\bibitem{Zeng96}
Y. B. Zeng and Y. S. Li, Acta Mathematica Sinica, New Series,
12, 217 (1996).

\bibitem{ma94}
W. X. Ma and W. Strampp, Phys. Lett. A 185, 277 (1994).

\bibitem{ma96}
W. X. Ma, B. Fuchssteiner and W. Oevel, Physica A 233, 331 (1996).

\bibitem{Zeng99}
Y. B. Zeng,
Physica A 262,405 (1999).

\bibitem{mel88}
V. K. Mel'nikov,
Phys. Lett. A, 133, 493 (1988).

\bibitem{mel92} 
V. K. Mel'nikov, 
Inverse Probl. 8, 133 (1992).


\bibitem{Zeng00}
Y. B. Zeng, W. X. Ma and R. L. Lin,  J. Math. Phys. 41(8), 5453 (2000).

\bibitem{salle}
V. B. Matveev and M. A. Salle,  Darboux transformations and solitons (Berlin: Springer, 1991).

\bibitem{zakharov}
V. E. Zakharov and A. B. Shabat,  Funk. Anal. Appl. 8, 226 (1974).

\bibitem{orlov}
A. Yu Orlov and S. R. Wojciechowski, 
Physica D, 69, 77 (1993).

\bibitem{chen}
Y. Chen and Mei A, Chinese Anal. A, 20, 667 (1999).


\bibitem{newell}
A. C. Newell, Solitons in Mathematics and Physics(Philadelphia: SIAM, 1985)

\bibitem{zeng91}
Y. B. Zeng, 
Phys. Lett. A 128, 488 (1991).



\end {thebibliography}
\end{document}